\documentclass[aps,prl,twocolumn,showpacs,superscriptaddress,
amsmath,amssymb,tightenlines]{revtex4}

\newcommand{\eq}[1]{Eq.~\eqref{#1}}
\newcommand{\half}{\hbox{$1\over2$}}
\newcommand{\<}{\langle}
\renewcommand{\>}{\rangle}

\usepackage{graphicx}

\begin{document}

\title{Nonlinear resonances in $\delta$-kicked Bose-Einstein condensates}
\author{T.S. Monteiro}
\email{t.monteiro@ucl.ac.uk}
\affiliation{Department of Physics and Astronomy,
University College London, Gower Street, London WC1E 6BT, United Kingdom}
\author{A. Rancon}
\affiliation{Department of Physics and Astronomy,
University College London, Gower Street, London WC1E 6BT, United Kingdom}
\author{J. Ruostekoski}
\affiliation{School of Mathematics, University of Southampton,
Southampton, SO17 1BJ, United Kingdom}

\date{\today}
\begin{abstract}
We investigate the Quantum Resonance (QR) regime of a periodically
kicked atomic Bose-Einstein condensate. We find that the clearest
indicator of the nonlinear dynamics
is a surprisingly abrupt cut-off which appears on the main ``Talbot time''
QR. We show that this is due and excitation path combining both
Beliaev and Landau processes, with some analogies to nonlinear self-trapping.
 Investigation of dynamical instability
reveals further symptoms of nonlinearity such as a regime of exponential oscillations.
\end{abstract}
\maketitle

The interaction between nonlinear dynamics and quantum dynamics has
stimulated new domains of cold atom
physics. For instance, the suppression of chaotic classical diffusion in a
corresponding quantum system
evolving under the {\em linear} Schr\"odinger equation has been investigated
experimentally using cold atoms
subjected to regularly spaced short pulses ($\delta$-kicks) from standing
waves. But studies of the quantum
$\delta$-kicked systems, which have as a classical analogue the famous
nonlinear paradigm which is
the Standard Map, had certain limitations: probing the dynamics close to
the  chaotic classical limit
requires small values of Planck's constant,
$\hbar$. Even though $\hbar$ in the cold atom experiments is a scaled (and
hence variable) effective value,
   the $\hbar \to 0$ regimes remained out of reach. However,  an important
advance was the realization in 2002
\cite{Fishman} that atoms kicked with periods which are a rational
multiple
of  the Talbot time $T=4 \pi$ (so-named because of an analogy to the
Talbot optical effect) represent
a somewhat pathological case. They correspond to a classical `image'
system where the effective value
  of  $\hbar$ is played by the dephasing from resonance and so can take
very small values. This has stimulated
a series of delicate experiments by several groups worldwide \cite{EXPT}
probing the rich physics of
both the Quantum Resonance (QR) regime
  as well as the closely related Quantum Accelerator (QA) regime (which is
in effect, quantum resonant
behavior in the presence of gravitational acceleration).  Further theory
\cite{THEO}
includes proposed applications such as the realization of a quantum random
walk algorithm \cite{Gardiner}.

Since well-defined initial momenta are needed, most recent
QR or QA experiments employed $\delta$-kicked  Bose-Einstein condensates (BECs).
But this in turn opens up a new and quite different possibility: namely
the largely unexplored
regime where nonlinear dynamics, arising from the many-body nature of the
BEC, combine
with the $\delta$-kicked quantum dynamics. To date, the deep understanding
of collective excitations,
acquired from other areas of BEC physics, has not been applied to the
unique dynamical features of
  the $\delta$-kicked atoms.

A few theoretical studies have considered interactions
  in QR/QA:  in ~\cite{Wimberger}  dephasing  of
resonant transport at the Talbot time
  as a function of nonlinearity parameter $g$ was found; in ~\cite{Zhang} exponential growth
of non-condensate atoms was
predicted for certain parameters at half the Talbot time; in
~\cite{Rebuzzini2} significant
  differences were reported for QA dynamics between attractive and
repulsive interactions.
Only recently, though, was it demonstrated \cite{Reslen} that an approach
based on Bogoliubov phonon modes is
essential; in particular, the onset of dynamical instability
was attributed to  parametric resonances. Parametric
instabilities through periodic driving
of collective modes have been investigated in several BEC
studies \cite{PR}.
Higher-order effects, resulting from phonon-phonon
interactions, have
been experimentally studied, e.g., in the context of excitation lifetimes
\cite{STA98} and
in a nonlinear coupling between two phonon modes \cite{HOD01}.
 Beliaev and Landau (BL) couplings provide the dominant contribution to
such nonlinear mode conversions; their importance was
 demonstrated in a series of recent experiments
with BECs excited by optical Bragg pulses \cite{Davidson}.

Here we have, for the first time, quantitatively investigated an
interacting $\delta$-kicked BEC by mapping the position and properties of
the resonances of all its low-lying modes.
The leading Talbot time
   resonances, which appear to dephase and disappear at $T=4\pi$, in
fact are shifted and broadened (they are strong over  $\Delta T \sim 1$).
But our key finding is that they acquire an extraordinarily sharp cut-off at their maximum ($\Delta T \sim
10^{-3}$).
  We show this behavior is due to a nonlinear feedback process, originating
from the resonant kicking and a combination of {\em both}
Beliaev and Landau coupling.
  In other nonlinear resonances, we find novel features,
  not previously seen in other driven BECs, such as
{\em exponential  oscillations} and Fano-like profiles.
  We calculate the local Lyapunov exponents and model the Fano profiles
quantitatively.
The kicked-BEC experiments to date, have used effective
  values of the  nonlinearity  ($g \sim 0.5$) which are only slightly
smaller than
  those ($g \gtrsim 1$) needed for the effects we find.

For non-interacting single-particle dynamics, an
 attractive features of
the $\delta$-kicked system, for quantum chaologists, is the
extensively-investigated
  simple quantum map which stroboscobically evolves the system from kick
$n$ to kick $n+1$.
For example, expressing our quantum state in a momentum basis, $\psi(x,t)=
\sum_{l=0}^\infty a_l(t)\ | l >$,  we write:
\begin{eqnarray}
{\bf a}((n+1)T)= {\bf U}_{g=0}(T) \ {\bf a}(t=nT)\,,
\label{map}
\end{eqnarray}
  where ${\bf a}(t=nT)$ is a vector with the amplitudes $a_l$.  The atoms
experience a kicking
potential $V_{kick}(x,t)= K \cos x \sum_n \delta (t-nT)$. The
corresponding unitary time evolution operator
  factors (exactly) into a free-evolution part $U_{free}$ and a kick part
$U_{kick}$, i.e.,
\begin{eqnarray}
{\bf U}_{g=0}(T)= U_{free}  \  U_{kick} = e^{-i \frac{\hat{L}^2 T}{2}} \
e^{-i K \cos x}\,.
\label{map1}
\end{eqnarray}
  The first exponential term represents the free evolution
under the kinetic energy operator
in some units where the atom mass $M=1$ and $\hbar=1$.
Clearly, since the atomic momentum $l=0,\pm 1,\pm 2...$ is
quantized in units of the recoil momentum,
  $T= 4\pi$  implies $ U_{free}=1$ for {\em all}
momentum states, so consecutive kicks add in phase.
The result is a phase-matched absorption
  of energy from the field yielding ballistic transport.  In contrast, the
non-resonantly kicked
atoms experience diffusive growth in energy.

We consider a uniform, tightly-confined, effectively 1D BEC with periodic boundary conditions.
 The corresponding field-free, many-body Hamiltonian,
 in a momentum representation, is:
\begin{eqnarray}
H= \sum_l \epsilon_l \hat{a}^{\dagger}_l   \hat{a}_l  +
      \frac{g_{1D}}{2L}\sum_{l,j,m}
        \hat{a}^{\dagger}_l  \hat{a}^{\dagger}_j \hat{a}_m  \hat{a}_{j+l-m}\,,
\label{MB}
\end{eqnarray}
where $\epsilon_k=\hbar^2 k^2/(2M)$ and $L$ is the BEC size. The 1D interaction constant
$g_{1D}\simeq 2\hbar a_s\omega_\perp$, depends on the atomic scattering
length $a_s$ and the transverse trap frequency $\omega_\perp$.
In the presence of linearized perturbations around the macroscopically occupied $k=0$ ground
state,  $H$  may be diagonalized up to quadratic order by the
Bogoliubov transformation
$\hat{a}_k = u_k \hat{b}_k - v_k \hat{b}_{-k}^\dagger$,
for $k\neq0$, with $u_k-v_k =1/(u_k+v_k) =\sqrt{\hbar\omega_k}$, where $\hbar\omega_k=
[ \epsilon_k (\epsilon_k+ 2g_{1D}n)]^{1/2} $ and $n=N/L$ is the atom density. Then we can expand $H$
in the orders of $\sqrt{N}$: $H = {\rm const}+ H^{(2)} +H^{(3)}+H^{(4)}$ where
$H^{(2)}= \sum_{k\neq0} {\hbar \omega_k} \hat{b}^{\dagger}_k   \hat{b}_k $ and
the cubic part $H^{(3)}$ describes the leading order contribution to the interactions between phonons
\begin{equation}
H^{(3)}=\kappa\sum_{q,p}( \Gamma_{qp}\hat{b}_{q}^\dagger \hat{b}_{p}^\dagger
\hat{b}_{-q-p}^\dagger+\Delta_{qp} \hat{b}_{p}^\dagger\hat{b}_{q}^\dagger\hat{b}_{q+p}+{\rm H.c.})\,,
\label{quartic}
\end{equation}
where $q,p,(q+p)\neq0$ and $\kappa={\sqrt{N} g_{1D}}/{L}$. The coefficient $\Gamma_{qp}$
 represents a process in which three phonons are created or annihilated and $\Delta_{qp}$ a
 process in which one phonon with momentum $q+p$ decays into two phonons with momenta
 $q$ and $p$ (Beliaev term) or its inverse in which two merge to produce a third (Landau term).
 Since the energy of the excitations around the ground state is positive, the processes described
 by $\Gamma_{qp}$ are suppressed by energy conservation. In
terms of the Bogoliubov amplitudes we obtain
\begin{align}
\Delta_{qp} &=u_p u_q u_{q+p}+2u_pv_qv_{q+p}-2u_qv_pu_{q+p}-v_pv_qv_{q+p}\nonumber\\
\Gamma_{qp} &=u_q v_p v_{q+p}-u_p u_q v_{q+p}
\label{coeff}
\end{align}
We assume macroscopic occupancy for the low-lying modes of interest and that phase fluctuations
 of the 1D condensate may be neglected. Hence we will treat the Bogoliubov mode amplitudes classically.
The $\delta$-kicked map (\ref{map1}), in the presence of the quadratic Hamiltonian $H^{(2)}$ alone,
 requires only a straightforward modification to its free-evolution part:
\begin{eqnarray}
{\bf U}_{g} (T) = {\cal B}^{-1} {\bf e^{-i  \omega T }} \ {\bf {\cal I}} \  { \cal B} \  U_{kick}
\label{map2}
\end{eqnarray}
Where $\cal B$ denotes the Bogoliubov transform, ${\bf \exp{(-i {\bf \omega} T)}} $
is a row vector with the mode frequencies and ${\bf{\cal I}}$ is the identity. The eigenvalues of
the non-unitary matrix ${\bf U}_{g} (T)$ indicate dynamical stability. It is easy to prove they
in general come in quartets  $\lambda, 1/\lambda,\lambda^*, 1/\lambda^{*}$. Then
$|\lambda_{max}| > 1$ (where $\lambda_{max} $ is the largest eigenvalue and the local Lyapunov exponent)
 imply dynamical instability and exponential growth in the relevant modes (at least for short times).

 We compare \eq{map2} to the numerical solutions of the 1D Gross-Pitaevski equation (GPE) (here
 rescaled to dimensionless units \cite{preprint}) for an initially uniform BEC:
\begin{equation}
i \partial_t \psi (x,t) = \left[- \half \partial_x^2
+ g | \psi(x,t) |^2 + K \cos x \ F_t\right]\psi(x,t)
\label{gpe}
\end{equation}
where $F_t= \sum_n  \delta(t - nT)$.
We study $g \simeq 0-10$, realistic with the current experiments \cite{Henderson,Davidson},
for $K <1$, which allows only the creation of discrete low-lying phonon excitations.

\begin{figure}[htb]
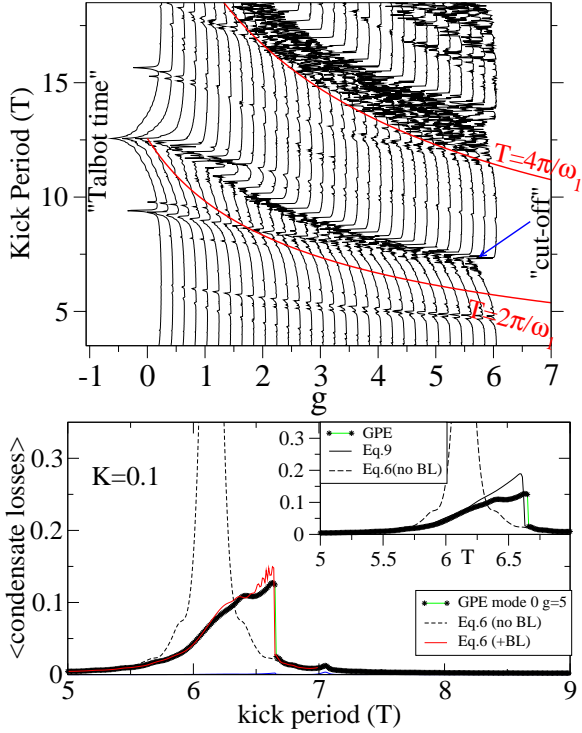

\includegraphics[width=3in]{Fig1a.eps}
\vspace*{10 mm}
\includegraphics[width=3.in]{Fig1b.eps}
\vspace*{-10 mm}
\caption{{\bf(a)} Losses from the unperturbed condensate $<1-|a_0(t)|^2>_t$ averaged over time,
calculated from the GPE, for kick strength $K=0.5$. The
BEC resonance evolves from the well-known Talbot time resonance of the QKR
(at $T=4\pi M$, $M=1,2\ldots$, for $g=0$) and  develops an abrupt cut-off for any $g \gtrsim 1$.
{\bf (b)}Illustrates the sharpness of the cut-off. GPE calculations with even 100 points (black stars)
 for every unit of $T$ still fails to resolve the cut-off.
 Plot for $g=5$, $K=0.1$. The quantum map {\em excluding} BL processes, incorrectly
gives a symmetric, unshifted resonance.
Inclusion of BL coupling perfectly reproduces not only the shift but also the cut-off; the simple
\eq{Simple} also provides reasonable agreement (inset).}
\label{Fig1}
\end{figure}

\begin{figure}[htb]
\includegraphics[width=3in]{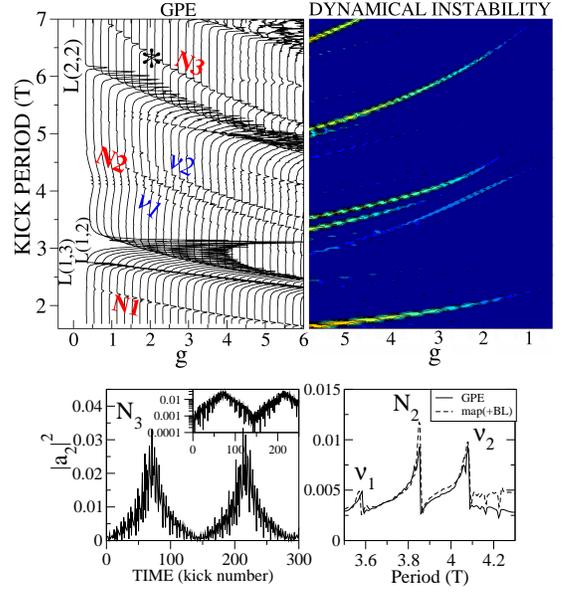}
\vspace*{10 mm}
\includegraphics[width=2.5in]{Fig2b.eps}
\vspace*{-10 mm}
\caption{Upper figure (left): Probability for mode 2 averaged over 100 kicks for $K=0.5$;
(right) Shows $|\lambda_{max}|$, largest eigenvalue of $U_{g}$. Bright regions
 denote $|\lambda| >1$ and hence exponential behavior (dynamical instability).
 The unstable $N_1,N_2,..N_n$ series of nonlinear resonances (which only appear for
$g \gtrsim 1 $) correspond to  $(\omega_1+ \omega_2) \simeq 2 n \pi T$.
 The asterisk denotes position of the `instability
border' found by \cite{Zhang}, which is thus due to $N_3$. The $L$ series are
 resonances which evolved from partial or full resonances of the Talbot
 time at $g=0$. They are stable, but in spite of this
 are {\em much} stronger than the exponential resonances.
Lower figure (left). Mode 2 probability of $N_3$, near the asterisk of upper figure. $T=6.12$, $g=2.5$.
The exponential growth persists for only a finite time; it is then replaced by exponential
decay, leading to {\em exponential oscillations}(log scale shown in inset) . Lower figure(right). Mode 2 near $N_2$. The cluster of
3 interacting resonances have ``Fano-like''  profiles. These are well reproduced
by \eq{map2} corrected with $H^{(3)}$ in \eq{quartic}, including only the 7 lowest modes.}
\label{Fig2}
\end{figure}

Fig.\ref{Fig1}(a) shows GPE numerics; it maps the average BEC response,
$<1-|a_0(t)|^2>_t=\frac{1}{t} \sum_{n=1}^{t} 1-|a_0(nT)|^2$ averaged over the first $t$ kicks, for $K=0.5$.
 At $g \simeq 0$, the Talbot time resonance at $T=4\pi$
is perfectly symmetrical, (as are the fractional resonances on either side).
For $g \simeq 0-1$ an asymmetry develops, due here to the lifting of the degeneracy between
the lowest modes: the main resonance splits into a mode
1 resonance $ \omega_1 T \simeq 2\pi$ and mode 2 resonance $ \omega_2 T \simeq 8\pi$.
The mode 2 resonance rapidly decays away as the gap  $\omega_2 -\omega_1$ increases:  direct
coupling $<0|U_{kick}|2> \sim J_2(K)$  with the condensate is small.
 A slight asymmetry was noted in the GPE numerics in Ref.~\cite{Wimberger} for the Talbot
time resonance at $g \simeq 0.1$, which we now attribute to this regime.

But the most striking feature is the very sharp `cut-off'
appearing at $ g \gtrsim 1$
(and still exists even at $g=20$).
 It is also evident in the second harmonic of the resonance (upper half of the graph).
Fig.\ref{Fig1}(b) shows that
 even a grid of 100 points per unit of T (each circle representing a
GPE simulation for $30$ kicks and $g=5$) is too coarse to resolve the cut-off
(in comparison, such a grid could  resolve the famously narrow $g=0$ Talbot-time resonance).

The dotted line shows the Bogoliubov map (\ref{map2}) here
 incorrectly produces a symmetric resonance and fails to shift the resonance
away from $\omega_1 T =2\pi$.
Hence we need to include the neglected phonon-phonon interaction terms between the
kicks from $H^{(3)}$.
The free-ringing ${\bf \exp{(-i  \omega} T )}$ part of the map [\eq{map2}] must be replaced
by a set coupled equations following from the Heisenberg equations
$d_t\hat{b}_k=-i\omega_k \hat{b}_k -i[\hat{b}_k,H^{(3)}/\hbar]$, for $k\neq0$, where we replace
$\hat{b}_k$ by the rescaled classical amplitudes $\tilde{b}_k\rightarrow \< \hat{b}_k\> /\sqrt{N}$. Figure~\ref{Fig1}(b) shows only four lowest excitations
($k=\pm 2 ,\pm 1$) provide an excellent quantitative agreement
with the GPE, accurately reproducing the cut-off for small depletion ($\lesssim 10\%$)
of the ground state.

We simplify further by transforming (by symmetry
$\tilde{b}_k=\tilde{b}_{-k}$) to the basis $ | l \rangle \to \frac{1}{\sqrt{2}} (|l \rangle + |-l \rangle)$,
for $l \neq 0$, so that
$\left< l | U_{kick}| n \right>= U_{nl}= i^{l-n}J_{n-l}(K) + i^{-(l+n)} J_{l+n}(K)$ if $n,l > 0$,
but $U_{0l}=\sqrt{2}i^{-l}J_l(K)$ and $U_{00}=J_0(K)$. We then
only need two coupled equations to accurately reproduce the cut-off. Moreover, neglecting the small $\Gamma_{qp}$ terms, we obtain (from the free-ringing plus BL terms)
\begin{align}
d_t \tilde{b}_1 &=-i [\omega_1  + 2C_1 Re(\tilde{b}_2) ]\tilde{b}_1 +2C_2 \tilde{b}_1^*\tilde{b}_2 , \nonumber \\
d_t \tilde{b}_2 &=-i\omega_2 \tilde{b}_2- i [C_1 |\tilde{b}_1|^2 +C_2 \tilde{b}_1^2 ]\,,
\label{pair}
\end{align}
where $C_1=\bar\kappa(\Delta_{-1,2}+\Delta_{2,-1})$, $C_2=\bar\kappa \Delta_{1,1}$, and
 $\bar\kappa=g/(2\sqrt{2}\pi)$ (note $\Delta_{-1,2}=\Delta_{1,-2}$, etc.).
 We can even obtain a reasonable cut-off if
we set the smaller term  $C_2=0$. Hence, the main effect of mode 2 is simply to
provide a phase-shift on $\omega_1$. If we
 integrate $\tilde{b}_2$  while keeping $|\tilde{b}_1|$ constant,
  we can replace the {\em full map} by:
\begin{equation}
d_t \tilde{b}_1 \simeq -i\omega_1 \tilde{b}_1 +i A \sin^2(\omega_2t/2)|\tilde{b}_1|^2
\tilde{b}_1 -i{\cal R}(t)\,,
\label{Simple}
\end{equation}
where $A=4C_1^2/\omega_2$, ${\cal R}(t)= \sqrt{2}J_1(K)(u_1-v_1) F(t)$.
Fig.\ref{Fig1}(b) inset shows that we still get  reasonable agreement with the full model
for weak $K =0.1$. In this regime $|\tilde{b}_0(t)|^2 \simeq 1-|\tilde{b}_1(t)|^2$.
Writing $\tilde{b}_1= \rho e^{i\theta}$, a phase space analysis in the $\rho,\theta$ plane
reveals a separatrix curve which appears at the cut-off parameters and bounds the value of $\rho$.
We may also describe the physical mechanism thus: if the $0 \to 1$ transition
is initially only slightly off-resonant, the kicking starts populating mode 1
effectively; with the increasing mode 1 population, BL processes begin
to populate the empty modes $l=\pm2$; this mode 2 population provides a phase-shift bringing
mode 1 closer into resonance; the nonlinear feedback accelerates the growth
in $b_1$ which in turns brings mode 1 further into resonance. However, if the $0 \to 1$ transition
 is initially too far off-resonant (beyond the sharp
cut-off value) the nonlinear feedback cycle cannot start and the population oscillations
between the modes are suppressed. An analogous model of two-mode dynamics
with continuous driving rather than $\delta$ kicks is reminiscent of a macroscopic
self-trapping effect in a BEC in a double-well potential \cite{SME97}, but in the
present work the cut-off is {\em considerably sharper}.

Fig.\ref{Fig2} maps the average probability of mode 2 (averaged over 100 kicks) for $K=0.5$. The right
hand side maps regions of dynamical instability $|\lambda_{max}| >1$.
We analyze dynamical stability by mapping the eigenvalues of ${\bf U}_{g} (T)$ for  all the resonances
of the lowest 3 excited modes. We divide the resonances into (1) the ``linear'' family $L(n,l)$ (ie those which
evolve from the linear case and converge at $g=0$ to a rational fraction of the Talbot time. The resonance in
Fig.\ref{Fig1}(a) is the $L(1,1)$ (first resonance of mode $l=1$). (2) The ``nonlinear'' resonances
$N_n$ and $\nu_n$ which vanish in the absence of interactions, at $g=0$; the $N_n$ correspond to
$(\omega_1+\omega_2) T \simeq 2\pi n$, while $\nu_n$ are somewhat analogous to ``counter-propagating mode'' resonances
found in modulated traps \cite{Dalfovo} and imply $2\omega_n T \simeq 2\pi$.
Contrary to the suggestion of \cite{Reslen} where no Liapunov exponents were calculated, we find that
none of the $L(n,l)$ resonances have any $|\lambda| >1$.
They are all stable, including $L(1,1)$, by far the strongest of all.
 But counter-intuitively, they
are associated with a much stronger BEC response, even after a very long-time (100 kicks)
than the nonlinear resonances $N_n$ and $\nu_n$ which are unstable.

 The reason for this is clear
from Fig.\ref{Fig2}(b). The mode 2 populations from the GPE (for both $N_1$ and $N_3$)
 grow exponentially for a finite time, then
decay exponentially; the inset shows this behavior on a log-scale. The map with BL
corrections here is quantitative for only the first 10-20 kicks, so we cannot model this behavior
 from $H^{(3)}$ alone.
 But it is tempting to attribute
it to regimes where either the $\lambda,\lambda^{*}$ or the $1/\lambda,1/\lambda^*$ eigenvalues
are predominant.
The cluster of interacting resonances $N_2,\nu_1$ and $\nu_2$ lies in a region of very low ground state
depletion (it lies in the mimimum of the dominant $L(1,1)$ tail).
 The map (\ref{map2}), corrected by BL terms $H^{(3)}$ in \eq{quartic}
 (with the lowest 7 modes),
reproduces very
well the characteristic Fano-like profiles seen in all three peaks, while the uncorrected map
 produces only symmetric resonance profiles.
To our knowledge neither the exponential oscillations, nor the Fano profiles are seen in
comparable non-equilibrium BEC dynamics. While not fully understood, they indicate that
the $\delta$-kicked systems offer new and experimentally accessible BEC dynamics.

While the Talbot time $g=0$ resonances have been proposed for metrological applications (e.g.,
for measurement of gravity), the similarly sharp BEC cut-off suggests analogous possibilities
as it provides a sharp excitation threshold. One could envisage a set of ring-BECs subject
to kicks; at threshold kicking frequencies, a very small perturbation would determine whether
a particular ring is left in an $l=0$ state or acquires a large $l=1$ component.
Moreover, even under a very weak rotation, a BEC in a large ring would occupy a higher momentum
(vortex) state $l$. Then the system could provide a sensitive probe of rotation for small
changes in the resonance frequency between different angular momentum states.

\newpage

\begin{thebibliography}{99}
\bibitem{preprint} We introduce dimensionless units $x= 2k_L x',t=8\omega_r t'$, where
the lattice recoil frequency $\omega_r\equiv\hbar k_L^2/2M$ and
 $g=\frac{Ng_{1D}}{(8\hbar\omega_r)} \frac{2\pi}{L}$.
Since $k_L L=\pi$, $L\rightarrow 2\pi$ and the BEC occupies $0\leq x\leq 2\pi$. Also $2gn \rightarrow \frac{gN}{\pi}$ and the Bogoliubov spectrum $\omega^2_{k}\rightarrow k^2/2( k^2/2+g/\pi)$.
\end{thebibliography}
\end{document}